# Thermal Activation of Dry Sliding Friction at The Nano-scale

Rasoul Kheiri[1, a)] and Alexey A Tsukanov[1]

[1] *Skolkovo Institute of Science and Technology, Moscow, 121205, Russia*

[a)] Corresponding author: Rasoul.Kheiri@skoltech.ru

**Abstract.** Molecular dynamic (MD) simulations are applied to investigate the dependency of the kinetic friction coefficient on the temperature at the nano-scale. The system is comprised of an aluminum spherical particle consisting of 32000 atoms in an FCC lattice sliding on a stack of several layers of graphene, and the simulations have done using LAMMPS. The interaction potential is charge-optimized many-body (COMB3) potential and a Langevin thermostat keep the system at a nearly constant temperature. With an assumption of linear viscous friction, $F_{fr} = -\gamma v$, the kinetic friction coefficient $\gamma$ is derived and plotted at different temperatures in the interval of T $\in$ [1, 600] K. As a result, by increasing temperature, the kinetic friction coefficient is decreased. Consequently, while the friction is assumed as a linear viscous model, the results are similar to the thermal activation in atomic-scale friction. That is, (1) by increasing sliding velocity friction force will be increased and (2) by increasing temperature, kinetic friction coefficient decreases.

## INTRODUCTION

Macroscopically, the kinetic friction force on solids is proportional to the normal load and independent of the sliding velocity which refers to the Coulomb-Amounts laws of friction. In contrast, nanoscale friction experiments with atomic force microscopy (AFM) show that friction is related to the speed of sliding objects. Generally, it has been observed that friction increases with the sliding velocity at the nanoscale. However, the mechanism of friction is complicated for the variety of dissipation sources and materials, including temperature, wear, chemical bonds on the interface, adhesion, etc. Specifically, more studies show a logarithmic dependency between friction and sliding velocity [1-5]. Yet, as an approximation in some ranges of velocities and applied forces, a linear velocity dependence of friction could be quite sensible. For instance, in some experiments for a thin film, dissipation is close to linear viscous friction [6].

In classical statistical mechanics, the kinetic energy of atoms is proportional to temperature. Hence, on the atomic scale, investigating the dependency of the kinetic friction on temperature is significant. In the nano-scale, experimental observations with atomic force microscopy (AFM) reveal that friction generally decreases by increasing temperature. The measurement performed from the cryogenic temperature to hundreds of Kelvin degree and the studied materials are including Silicon, graphene, diamond, and other materials [7-13].[1]

Besides, using molecular dynamics (MD) simulations, as an alternative to AFM experiments, people understand better the dependency of friction on the sliding velocity, normal load, temperature, and other involving parameters [14] . Particularly, MD simulations help us to understand the interactions on the interface between the sliding object and the substrate, which is buried in the AFM experiments. For thermal effects, molecular dynamic (MD) simulations have reported the same dependency of decreasing kinetic friction with increasing temperature [7,15-18].

---

[1] Besides the general increase of friction by decreasing temperature, some research shows a peak around 50-200 K [10-13].

Theoretically, the mechanism is referred to as thermal activation. In particular, Prandtl-Tomlinson (PT) model [19, 20] has been utilized to interpret the experimental results for the dependency of friction on temperature [21]. To point out, considering a single particle and a potential barrier, the thermal energy facilitates the sliding of the particle from potential barriers. Therefore, for an AFM tip, the sliding is easier at a higher temperature and the tip can jump over potential barriers. As an example, a thermal simulation based on Prandtl-Tomlinson (PT) model is discussed in [22].

In a former paper [23], we introduced molecular dynamics (MD) simulations for a spherical metal nanoparticle sliding on several graphene layers. In this case, we showed that MD simulations for a thermally fluctuating surface at $T=300$ $K$ are in agreement with a linear viscous model of friction. That is, the friction force is $F_{fr} = -\gamma v_{st}$, in the range of the lateral and normal forces, $F_x \in [2, 5]$ $nN$, $F_z \in [15, 60]$ $nN$, respectively, exerted on the sliding nanoparticle, where $v_{st}$ stands for the steady state velocity of the sliding nanoparticle and $\gamma$ is the friction coefficient. In the current paper, firstly we perform similar simulations in different temperatures and derive the friction coefficient from $F_{fr} = -\gamma v_{st}$. In the second step, the friction coefficients will be plotted as a function of temperature.

## MOLECULAR DYNAMICS SIMULATIONS

### Simulation setup

Figure 1 shows snapshots of MD simulations in three temperatures; $T = 1$ $K$ (Fig. 1a), $T= 300$ $K$ (Fig. 1b), and $T= 600$ $K$ (Fig. 1c). The substrate is built upon graphene layers, and the tip is made up of aluminum atoms in the Face-Centered Cubic (FCC) lattice. The system firstly is equilibrated at each temperature while a vertical force $F_z$ is exerted on it. The equilibration steps are fixed box relax, minimization, and isothermal-isobaric (npt) barostat in LAMMPS [24]. The "fix npt command" is sampling positions and velocities from an isothermal-isobaric ensemble by the Nose-hoover style each time step [25].

After equilibration, a lateral force $F_x$ is added to the spherical particle. Then, the particle starts sliding in the x-direction and reaches a steady state regime in a few nanoseconds. However, sliding brings about some heat that needs to be removed from the system. Therefore, we add a Langevin thermostat to the system before time integration in each time step. In this way, the system fluctuates gently around a constant temperature. Particularly, the Langevin thermostat has a damping parameter as the inverse effect of viscosity, which is suitable for our assumption of linear viscous friction.

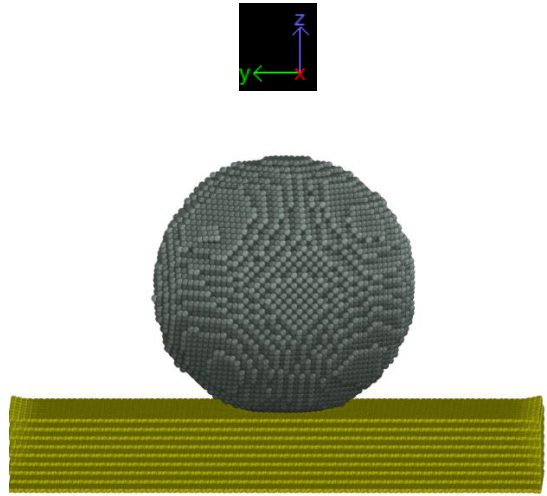

(a)

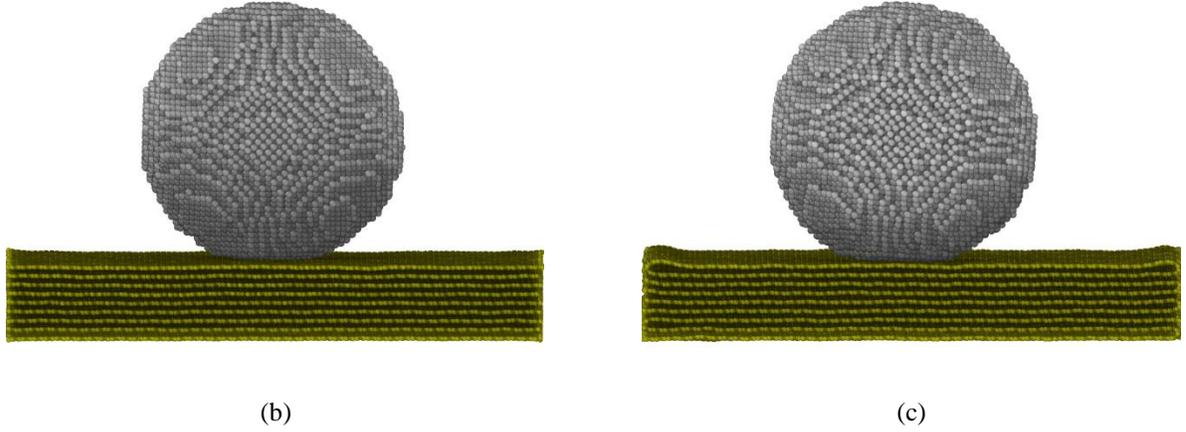

(b)                 (c)

**FIGURE 1.** Snapshots of the MD simulations in different temperatures (a) T = 1 K, (b) T = 300 K, and (c) T =600 K; A spherical aluminum particle is sliding in the x-direction on the graphene layers. The bottom graphene layer is frozen and the edges are fixed to prevent the surface from vertical and horizontal movement, respectively.

## Potential

Figure 2 shows that the potential for sliding friction could be complicated, especially if when the substrate and the tip are not from an identical material. Generally speaking, in the solid atomic scale, the resistance in sliding friction comes from the interaction between the sliding particle and the surface, or otherwise comes from surface properties like the surface asperities of lattice structure. For MD simulations, there is a cross-potential between the substrate and the sliding particle. As described in the first paper [23], the substrate consists of several graphene layers, and the sliding particle is made up of aluminum atoms. On one hand, for metal Al particles, Embedded Atom Method (EAM) is suitable [26]. For graphene layers, on the other hand, adaptive intermolecular reactive empirical bond-order (AIREBO) potential is appropriate [27]. However, the cross-potential or the tip-substrate potential as shown in Fig. 2 could be challenging since the interface interactions includes also Al-C. To have all potentials of C-C (in layer and between layers), Al-Al, and Al-C together, we have used third-generation charge optimized many-body (COMB3) potential, where 3 refers to the third generation of the COMB potential [28]. More description of COMB potential can be found in the reference [23]. We have used COMB3 potential as developed and applied for Al-C interactions by Sinnott, et al. [29]. The simulations were performed at different temperatures from 1 K to 600 K, and we didn't introduce any defeats in the substrate. Thereby, the interaction between aluminum atoms and graphene layers remains weak, and aluminum-carbic might not form in these conditions [29].

## Chiral index

In atomic-scale friction, the sliding orientation can change the friction force due to the anisotropic directions of lattices [30]. For example, sliding on commensurate and incommensurate angles can result in different amounts of friction or even structural lubricity [31-34]. In our case study, there is a metal FCC particle sliding on graphene layers. The orientation of sliding on graphene can be described by a chiral vector. That is $C = na_1 + ma_2$, where $a_1$, and $a_2$ are basis vectors of a graphene sheet. Two special case of chiral index are called Armchair ($n=0$), and Zigzag ($m=0$) directions [35, 36]. In the current simulations, the spherical aluminum particle is sliding in the Armchair direction of the substrate.

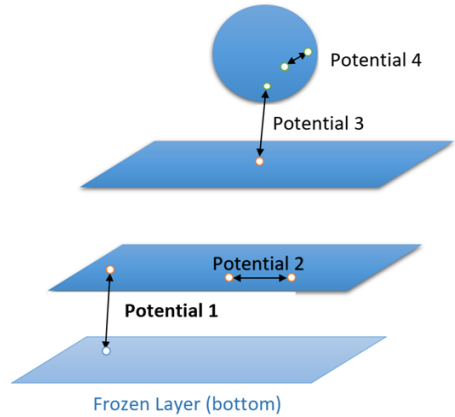

**FIGURE 2.** An illustration of the complexity of the potential. Cross-potential is required since the sliding particle (metal) and the substrate (graphene layers) are not from the same material.

# RESULTS

By applying a constant lateral force $F_x$ as well as the normal load $F_z$, the spherical particle illustrated in Fig. 1 starts to accelerate on the surface. Due to the existence of a velocity-dependent friction force, the acceleration reduces, and finally, the particle will reach a steady state regime, which takes a few nanoseconds. Afterward, the particle slides with a constant velocity, wherein the friction force $F_{fr}$ equals the lateral force $F_x$.

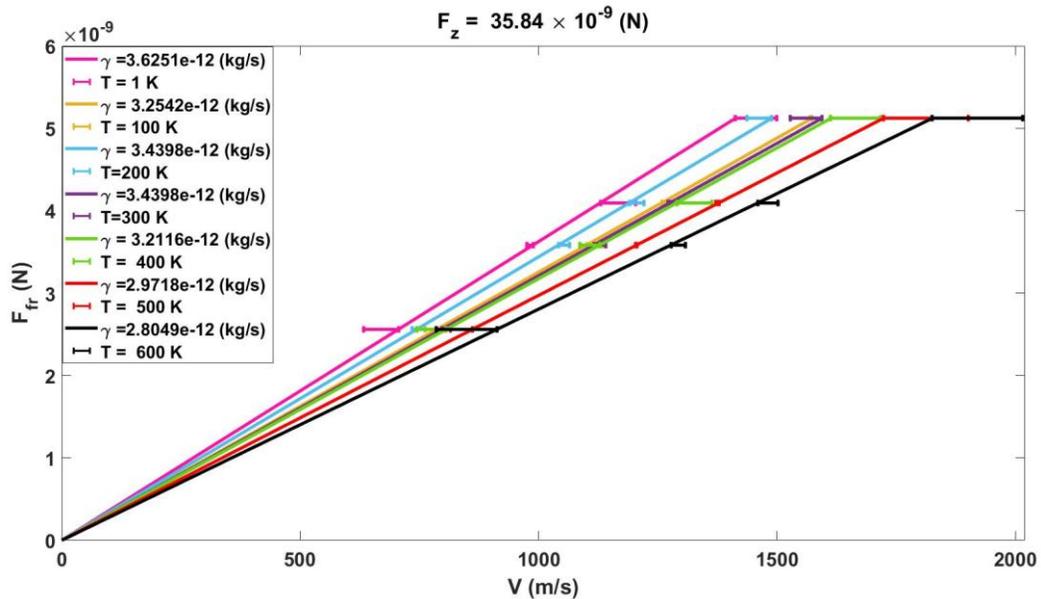

**FIGURE 3.** Velocity dependence of friction force for different temperatures at the constant normal load of $F_z = 35.84 \times 10^{-9} \, (N)$

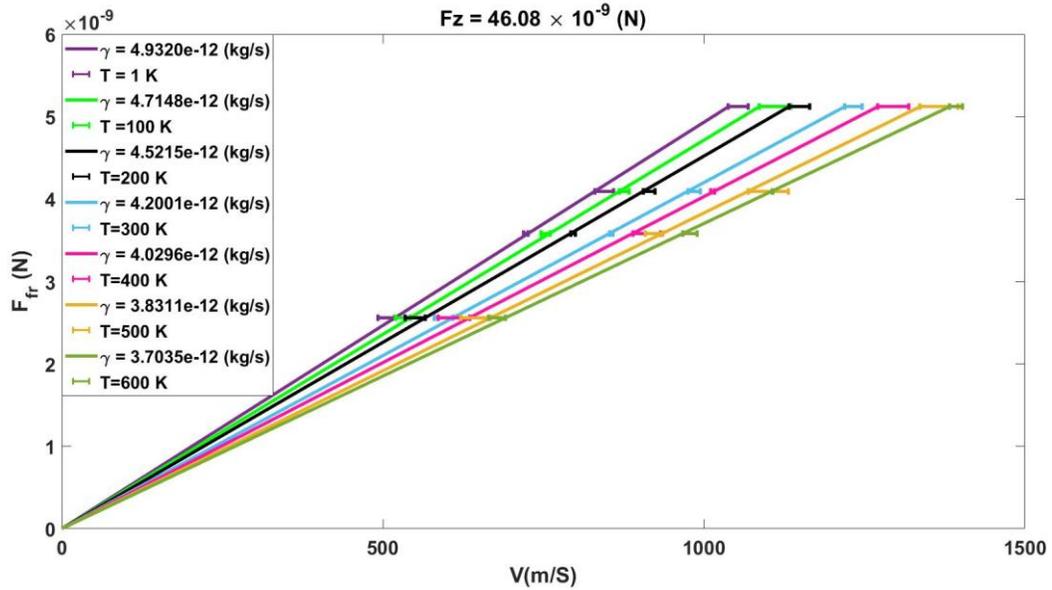

**FIGURE 4.** Velocity dependence of friction force for different temperatures at the constant normal of $F_z = 46.08 \times 10^{-9}$ (N)

At a steady state regime, we model the friction force to be approximate as a linear viscous dissipation $F_{fr} = -\gamma v_{st}$. This model then investigated different temperatures within a constant normal load. Then, we changed the normal load and repeated the simulations. Fig. 3 and Fig. 4 illustrate a linear dependency of friction force for a range of different temperatures $T \in [1, 600]$ K. The normal loads are $F_z = 35.84 \times 10^{-9}$ (N), and $F_z = 46.08 \times 10^{-9}$ (N) in Fig. 3, and Fig. 4, respectively. Moreover, the range of horizontal forces is $F_x \in [2, 5]$ nN. The slope of the lines in Fig. 3, and Fig. 4 are the kinetic friction coefficient for a linear viscous friction force versus the sliding steady-state velocity. These two figures then imply that as temperature increases, the friction coefficient decreases which is similar to the thermal activation theories and experimental results by AFM. This effect is illustrated in Fig. 4 explicitly.

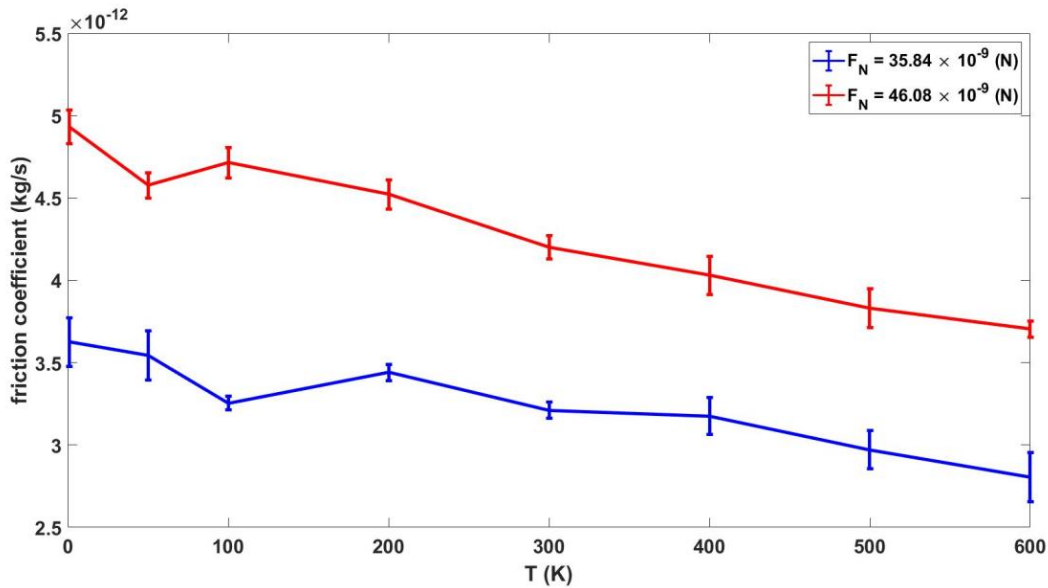

**FIGURE 5.** Friction coefficient as a function of temperature for two different applied normal load

As the main result, friction coefficient in terms of temperature is depicted in Fig. 5 for two different applied normal loads. Fig. 5 shows that friction coefficient is decreased with increasing temperature, that is similar to thermal activation models. Furthermore, it is shown that a higher normal load will add to the friction coefficient.

Lastly, we mention some other thermal observations, generally and shortly. We observed that at lower temperatures the spherical aluminum particle could show more solidity and deform more elastically by increasing the applied normal load. Moreover, in low temperatures, the particle could reach a steady state velocity more gently and faster. Furthermore, the stick-slip phenomenon for larger normal loads and lower temperatures could occur more clearly than higher temperatures and fewer normal loads.

## CONCLUSION

using LAMMPS, MD simulations were performed for a system of sliding nanoparticle on a stack of several graphene layers in different temperatures. The spherical nanoparticle consists of about 32000 aluminums in FCC lattice interacting weakly by the surface using COMB3 potential. The simulation time is about a few nanoseconds, which means that within a few nanoseconds, the sliding nanoparticle reaches a steady state velocity $v_{st}$. Moreover, during the simulation time, a Langevin thermostat is keeping the system in a nearly constant temperature.

In Fig. 2, and Fig.3, the velocity dependence of friction force is illustrated for different temperatures. Considering a linear viscus model, $F_{fr} = -\gamma v_{st}$, for the range of applied horizontal $F_x$ and vertical $F_z$ forces, these two figures are implying that as the temperature increases the friction coefficient $\gamma$ decreases. This is explicitly can be seen from Fig. 5, in consistency with the thermal activation theory. The investigation has done for temperatures $T \in [1, 600]$ K, and two different normal loads $F_z = 35.84 \times 10^{-9}$ (N), and $F_z = 46.08 \times 10^{-9}$ (N).

The simulations performed without charges. Yet, since COMB potential is a charge-inclusion potential, it can be utilized further to add the electronic contribution of friction and make a comparison between results of the electronic and phononic portion of kinetic friction coefficient.

We emphasize that while in the thermal activation theory, the dependency of friction force on velocity is logarithmic, we illustrated that a linear viscose friction shows the similar effect of thermal activation. That is, in either case, the kinetic friction force is increasing with velocity and by increasing temperature, friction decreases.

The simulations performed without charges. Yet, since COMB potential is a charge-inclusion potential, it can be utilized further to add the electronic contribution of friction and make a comparison between results of the electronic and phononic portion of kinetic friction coefficient. Furthermore, during MD simulations, the sliding velocity is typically higher than the AFM tip experiments [14], and thermal activation theory might not be always true in MD simulations with larger velocities [1].

## ACKNOWLEDGMENTS

The research was carried out using the HPC computing resource at Skoltech supercomputer Zhores [37]. The authors gratefully acknowledge Prof. NV Brilliantov for his recommendations and advice. RKh is thankful for communication with Prof. SB Sinnott, and Prof. AF Fonseca (https://research.matse.psu.edu/sinnott/research). RKh also acknowledges Prof. AM Vishnyakov, Ilia Kopanichuk, and EN Hahn (https://www.ericnhahn.com/) for our discussions.